\documentclass[aps,pra,twocolumn,showpacs,showkeys]{revtex4}
\usepackage{graphicx,amsmath,amssymb,float,bm,times}
\usepackage[breaklinks]{hyperref}

\begin{document}

\title{\bf\large{Quark and Gluon Condensates in Isospin Matter}}
\author{Lianyi He, Yin Jiang and Pengfei Zhuang}
\affiliation{Physics Department,
Tsinghua University, Beijing 100084, China}

\begin{abstract}
Applying the Hellmann-Feynman theorem to a charged pion gas, the
quark and gluon condensates at low isospin density are determined
by precise pion properties. At intermediate density around $
f_\pi^2m_\pi$, from both the estimation for the dilute pion gas
and the calculation with Nambu--Jona-Lasinio model, the quark
condensate is strongly and monotonously suppressed, while the
gluon condensate is enhanced and can be larger than its vacuum
value. This unusual behavior of the gluon condensate is universal
for Bose condensed matter of mesons. Our results can be tested by
lattice calculations at finite isospin density.
\end{abstract}

\date{\today}
 \maketitle

While the vacuum structure of Quantum Chromodynamics (QCD) is
rather complicated, quark and gluon condensates may be sufficient
to determine the spectral properties of many hadrons. The lowest
dimensional condensates in vacuum are~\cite{cohen}
$\langle\bar{q}q\rangle_0\simeq-(250\ \text{MeV})^3$ and $\langle
G^2\rangle_0\simeq(360\ \text{MeV})^4$ with the definition $
G^2\equiv\alpha_s/\pi G_{\mu\nu}^aG_a^{\mu\nu}$, where $\alpha_s$
is the QCD coupling, $q$ the light quark field, and $G^a_{\mu\nu}$
the gluon field tensor. On the other hand, the in-medium behavior
of the condensates are of great importance for us to understand
how the hot and dense environment modifies the vacuum structure
and the hadron properties. Especially, the quark condensate is the
order parameter of chiral symmetry restoration, and the gluon
condensate may be related to the deconfinement phase
transition~\cite{gluon}.

For systems at zero temperature but finite baryon density
$\rho_{\text B}$, the ratio $R_q(\rho_{\text
B})\equiv\langle\bar{q}q\rangle_{\rho_{\text
B}}/\langle\bar{q}q\rangle_0$ and the difference $D_g(\rho_{\text
B})\equiv\langle G^2\rangle_{\rho_{\text B}}-\langle G^2\rangle_0$
between the condensates in baryon matter and in vacuum can be
expressed as~\cite{cohen}
\begin{eqnarray}
\label{baryon}
R_q(\rho_{\text B})&=&1-\frac{\sigma_{\text
N}}{f_\pi^2m_\pi^2}\rho_{\text B}+\cdots,\nonumber\\
D_g(\rho_{\text B})&=&-{8\over 9}\left(m_{\text N}-\sigma_{\text
N}-S\right)\rho_{\text B}+\cdots,
\end{eqnarray}
where $m_\pi,\ f_\pi,\ \sigma_{\text{N}}$ and $S$ are pion mass,
pion decay constant, nucleon $\sigma$ term and strangeness content
of nucleon in vacuum, and $\cdots$ denotes the higher order
correction. Taking only the linear terms of $R_q$ and $D_g$ which
are valid at low density, the quark condensate at nuclear
saturation density is $25-50\%$ smaller than its vacuum value, but
the gluon condensate is reduced by only $3-6\%$. The large
uncertainty is from the $\sigma$ term which is not yet precisely
determined. To obtain the behavior of the condensates at high
density, effective models may be used.

Recently, the study on QCD in medium is extended to finite isospin
and strangeness densities\cite{lattice,cpt1}. The physical
motivation to discuss isospin matter is to understand the mechanism
of QCD phase transitions at finite density. While there is not yet
precise lattice result at finite baryon density due to the fermion
sign problem, it is in principle no problem to do lattice simulation
at finite isospin density~\cite{lattice}. The QCD phase structure in
isospin matter is also investigated in many low energy effective
models, such as chiral perturbation theory~\cite{cpt1,cpt2}, ladder
QCD~\cite{ladder}, random matrix method~\cite{random}, strong
coupling lattice QCD~\cite{sqcd}, and Nambu--Jona-Lasinio (NJL)
model~\cite{njl1,njl2,njl3,njl4}. Very recently, the isospin matter
is discussed in the frame of AdS/CFT~\cite{adscft}. In this paper,
we study the quark and gluon condensates in a model independent way
at low isospin density and compare the prediction with a model
calculation at high isospin density. \emph{While it is generally
believed that the density effect will reduce the condensates, we
will see that the gluon condensate at finite isospin density can be
enhanced and even be larger than its vacuum value. Our result is
qualitatively different from the case at finite baryon density, and
the physical reason may be the repulsive meson-meson interactions.}

Since pions are the lightest excitations of QCD carrying isospin,
the ground state of QCD at small isospin density can be considered
as a dilute gas of charged pions. Without loss of generality, the
matter is assumed to be composed of $\pi^+$ mesons with isospin
density $\rho_\text{I}>0$. The relevant $\pi-\pi$ interaction in
such a system is characterized by the s-wave $\pi-\pi$ scattering
length $a$ in $I=2$ channel. The value of $a$ is predicted many
years ago~\cite{pipi} and can be taken as $a=m_\pi/(16\pi
f_\pi^2)\simeq0.043m_\pi^{-1}$.

At zero temperature and low isospin density, when the condition
$\rho_{\text I}a^3 \ll 1$ is satisfied, the pion matter which is
in Bose-Einstein condensation state can be described by the LHY
(Lee, Huang and Yang) theory~\cite{lhy} which is accepted as a
general theory for weakly interacting Bose gas~\cite{anderson}.
The energy density of the system is written as
\begin{equation}
\label{lhy1}
{\cal E}=m_\pi\rho_{\text I}+\frac{2\pi a\rho_{\text
I}^2}{m_\pi}\left(1+\frac{128}{15\sqrt{\pi}}\sqrt{\rho_{\text
I}a^3}+\cdots\right).
\end{equation}
The first term on the right hand side corresponds to the rest
energy, and the second one is the interacting energy in LHY. Note
that the leading order correction $2\pi a\rho_{\text I}^2/m_\pi$
was predicted by Bogoliubov~\cite{bog} before LHY. The pressure
${\cal P}$ and isospin chemical potential $\mu_{\text I}$ can be
easily obtained from ${\cal E}$,
\begin{eqnarray}
\label{lhy2}
{\cal P}&=&\frac{2\pi a\rho_{\text
I}^2}{m_\pi}\left(1+\frac{64}{15\sqrt{\pi}}\sqrt{\rho_{\text
I}a^3}+\cdots\right),\nonumber\\
\mu_{\text I}&=&m_\pi+\frac{4\pi a\rho_{\text
I}}{m_\pi}\left(1+\frac{32}{3\sqrt{\pi}}\sqrt{\rho_{\text
I}a^3}+\cdots\right).
\end{eqnarray}
From these relations, one can determine the pion properties with
the equation of state of the system,
\begin{equation}
\label{limit}
m_\pi = \lim_{\rho_{\text I}\to 0}\mu_{\text
I}(\rho_{\text I}),\ \ \ \ a = \lim_{\rho_{\text I}\rightarrow
0}\frac{m_\pi {\cal P(\rho_{\text I})}}{2\pi\rho_{\text I}^2}.
\end{equation}

The behavior of the in-medium quark and gluon condensates at low
density can be derived in a model independent way. In the QCD
Hamiltonian density ${\cal H}_{\text{QCD}}$, chiral symmetry is
explicitly broken by the current quark mass term ${\cal
H}_{\text{mass}} = 2m_q\bar{q}q+m_s\bar{s}s+\cdots$, where $s$ is
the strange quark field, $m_q$ and $m_s$ are light and strange
quark masses, and $\cdots$ denotes heavy quark (c,b,t)
contribution which is irrelevant to our discussion. Possible
isospin breaking effect is neglected here, since it will not
change our result. According to the Hellmann-Feynman theorem of
quantum mechanics~\cite{qm}, we obtain
\begin{equation}
\label{hfqcd}
2m_q\langle\psi|\int d^3x\ \bar qq |\psi\rangle
=m_q\frac{d}{dm_q}\langle\psi|\int d^3 x{\cal
H}_{\text{QCD}}|\psi\rangle,
\end{equation}
where $|\psi(m_q)\rangle$ is the ground state of the system as a
function of light quark mass, and we have multiplied the equation
by $m_q$ to obtain renormalization-group invariant quantities.

Applying the Hellmann-Feynman theorem (\ref{hfqcd}) to the isospin
matter and to vacuum and taking into account the uniformity of the
system, we have
\begin{equation}
\label{qc}
2m_q\left(\langle\bar{q}q\rangle_{\rho_{\text
I}}-\langle\bar{q}q\rangle_0\right)=m_q\frac{d{\cal E}}{dm_q},
\end{equation}
where the derivative $d{\cal E}/dm_q$ is taken at fixed density.

The gluon condensate can be obtained by considering the trace of
the energy-momentum tensor $T_\mu^\mu=-9/8
G^2+2m_q\bar{q}q+m_s\bar{s}s$, where we considered only the $u,\
d$ and $s$ quarks and neglected the heavy quark contribution. The
difference between the expectation values of the trace of the
energy-momentum tensor in isospin matter and in vacuum is $\langle
T_\mu^\mu\rangle_{\rho_{\text I}}-\langle T_\mu^\mu\rangle_0={\cal
E}-3{\cal P}$, which leads to the following result for the change
in the gluon condensate,
\begin{equation}
\label{gc} \langle G^2\rangle_{\rho_{\text I}}-\langle
G^2\rangle_0 =-\frac{8}{9}\left[{\cal E}-3{\cal
P}-2m_q\left(\langle\bar{q}q\rangle_{\rho_{\text
I}}-\langle\bar{q}q\rangle_0\right)\right],
\end{equation}
where we have ignored the strangeness content of pions,
$m_s\left(\langle\bar{s}s\rangle_{\rho_{\text
I}}-\langle\bar{s}s\rangle_0\right)=0$.

Taking the Weinberg result $a\propto m_\pi/f_\pi^2$~\cite{pipi}
and neglecting the $m_q$-dependence of the decay constant $f_\pi$,
we have $da/dm_\pi=a/m_\pi$. Since the energy density ${\cal E}$
is only a function of $m_\pi$ at fixed $\rho_{\text I}$, the
derivative in (\ref{qc}) can be expressed as $d{\cal
E}/dm_q=(d{\cal E}/dm_\pi)(dm_\pi/dm_q)$. Combining with the
Gellmann-Oakes-Renner relation
$2m_q\langle{\bar{q}q}\rangle_0=-m_\pi^2f_\pi^2$ and the LHY
energy density (\ref{lhy1}), we finally obtain the ratio
$R_q(\rho_{\text I})\equiv\langle\bar{q}q\rangle_{\rho_{\text
I}}/\langle\bar{q}q\rangle_0$ and difference $D_g(\rho_{\text
I})\equiv\langle G^2\rangle_{\rho_{\text I}}-\langle G^2\rangle_0$
between the condensates in isospin matter and in vacuum,
\begin{eqnarray}
\label{qgc2}
R_q(\rho_{\text I})&=&1-\frac{\rho_{\text
I}}{2f_\pi^2m_\pi}-\frac{64\sqrt{\pi}
a^{5/2}}{5f_\pi^2m_\pi^3}\rho_{\text I}^{5/2}+\cdots,\\
D_g(\rho_{\text I})&=&-\frac{8}{9}\left(\frac{m_\pi}{2}\rho_{\text
I}-\frac{4\pi a}{m_\pi}\rho_{\text
I}^2-\frac{64\sqrt{\pi}a^{5/2}}{3m_\pi}\rho_{\text
I}^{5/2}+\cdots\right).\nonumber
\end{eqnarray}

The linear $\rho_{\text I}$-dependence of $R_q$ and $D_g$ can be
completely determined via only two parameters, the pion mass
$m_\pi$ and decay constant $f_\pi$. From the quark condensate, it
is natural to define a density scale $\rho_0=f_\pi^2m_\pi$, which
is approximately equal to the nuclear saturation density
$\rho_{\text{sat}}$ ($\rho_0\simeq1.1\rho_{\text{sat}}$), and the
linear dependence is valid for $\rho_{\text I}\ll \rho_0$. The
higher order correction in isospin matter is quite different from
the one in baryon matter. For the quark condensate, the leading
order correction in isospin matter is $O(\rho_{\text I}^{5/2})$,
while it is $O(\rho_{\text B}^{4/3})$~\cite{cohen} in baryon
matter due to its fermionic nature. For the gluon condensate,
since the coefficient $m_\pi/2$ of the linear term in isospin
matter is much less than the one ($m_{\text N}-\sigma_{\text
N}-S$) in baryon matter and the $\pi-\pi$ interaction is
repulsive, the competition between the linear and leading terms
may make the condensate to increase at high isospin density.
Neglecting the term $O(\rho_{\text I}^{5/2})$ which is indeed
small even at high density and using the Weinberg result
$a=m_\pi/(16\pi f_\pi^2)$~\cite{pipi}, the gluon condensate starts
to increase at $\rho_{\text I}=\rho_0$ and becomes larger than its
vacuum value at $\rho_{\text I}>2\rho_0$. While this prediction is
beyond the validity density region $\rho_{\text I}\ll\rho_0$ of
the LHY equation of state, we do expect that the gluon condensate
may go beyond its vacuum value at high isospin density. We will
examine it in the following with a chiral quark model.

To estimate the behavior of the quark and gluon condensates at
high density is beyond the above approach, since the composite
nature of pions may become important at high isospin density.
Especially, at extremely high $\rho_{\text I}$, the isospin matter
is expected to become a weakly coupled Fermi
superfluid~\cite{lattice}. In this case, the element constitutes
of the system are no longer pions but quarks. To have a complete
understanding of the quark and gluon condensates at finite isospin
density, we adopt an effective chiral model at quark level to
describe the evolution from a weakly interacting Bose condensate
to a Fermi superfluid.

One of the models that enables us to see directly how the dynamic
mechanism of chiral symmetry breaking and restoration operate is
the NJL model~\cite{njl0} applied to quarks~\cite{njlquark}.
Recently, this model is extended to finite isospin chemical
potential~\cite{njl1,njl2,njl3,njl4}. The Lagrangian density of
the model is
\begin{equation}
\label{njl}
{\cal L} =
\bar{\psi}\left(i\gamma^{\mu}\partial_{\mu}-m_q\right)\psi
+G\left[\left(\bar{\psi}\psi\right)^2+\left(\bar{\psi}i\gamma_5\mbox{\boldmath{$\tau$}}\psi\right)^2
\right],
\end{equation}
where $\psi=(u,d)$ is the two-flavor quark field. The current
quark mass $m_q$, the coupling constant $G$ and a high momentum
cutoff $\Lambda$ due to the non-renormalization of the model are
phenomenological parameters and can be determined by fitting the
pion mass, pion decay constant and quark condensate in vacuum. The
isospin density enters the model via introducing an isospin
chemical potential $\mu_{\text I}$, corresponding to the isospin
charge $I=\int d^3x\bar{\psi}\gamma_0\tau_3\psi$. The order
parameters for chiral symmetry breaking and isospin symmetry
breaking are respectively the quark condensate $\langle
\bar{q}q\rangle$ and pion condensate $\langle\bar{u}i\gamma_5
d\rangle$.

At mean field level, the thermodynamic potential of the isospin
matter in the NJL model can be evaluated as~\cite{njl4}
\begin{eqnarray}
\label{omega}
\Omega &=& -6\int {d^3 {\bf k}\over (2\pi)^3}\left(E_++E_--2\sqrt{{\bf k}^2+M_0^2}\right)\nonumber\\
&&+\frac{(M-m_q)^2-(M_0-m_q)^2+\Delta^2}{4G}
\end{eqnarray}
with the dispersions $E_\pm = \sqrt{(\xi_\pm)^2+\Delta^2}$,
$\xi_\pm=\epsilon\pm \mu_{\text I}/2$ and $\epsilon = \sqrt{{\bf
k}^2+M^2}$, where $M=m_q-4G\langle\bar{q}q\rangle$ is the in-medium
quark mass, $M_0$ is its vacuum value, and
$\Delta=-4G\langle\bar{u}i\gamma_5 d\rangle$ is a BCS-like energy
gap. The physical quark and pion condensates correspond to the
minimum of the thermodynamic potential,
\begin{equation}
\label{gap} {\partial\Omega\over\partial M}=0,\ \ \ \
{\partial\Omega\over\partial \Delta}=0,
\end{equation}
which together with the isospin density
\begin{equation}
\label{density}
\rho_{\text
I}=-{\partial\Omega\over\partial\mu_{\text I}}
\end{equation}
determine self-consistently $M, \Delta$ and $\mu_{\text I}$ as
functions of $\rho_{\text I}$.

\begin{figure}[!htb]
\begin{center}
\includegraphics[width=7cm]{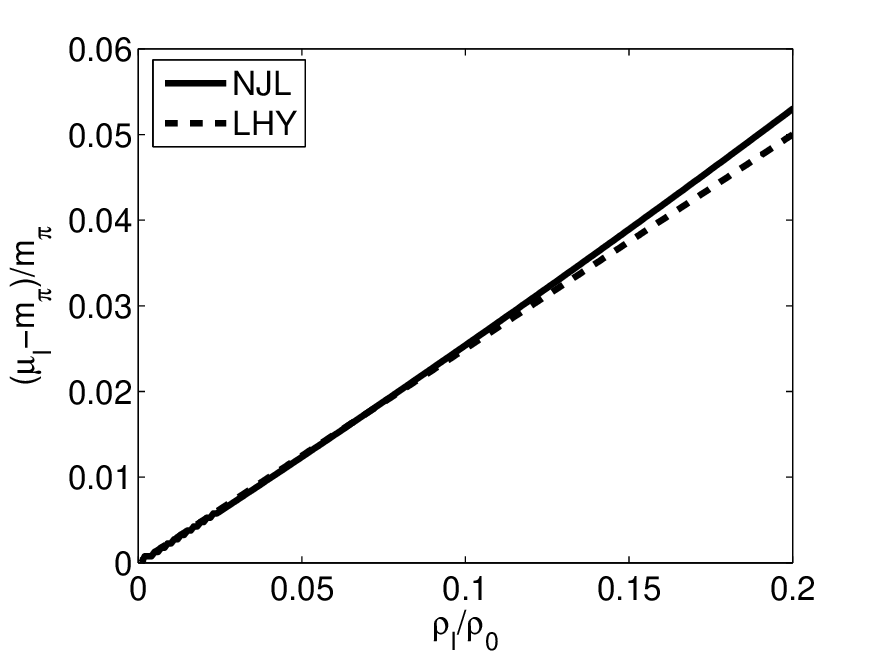}
\includegraphics[width=7cm]{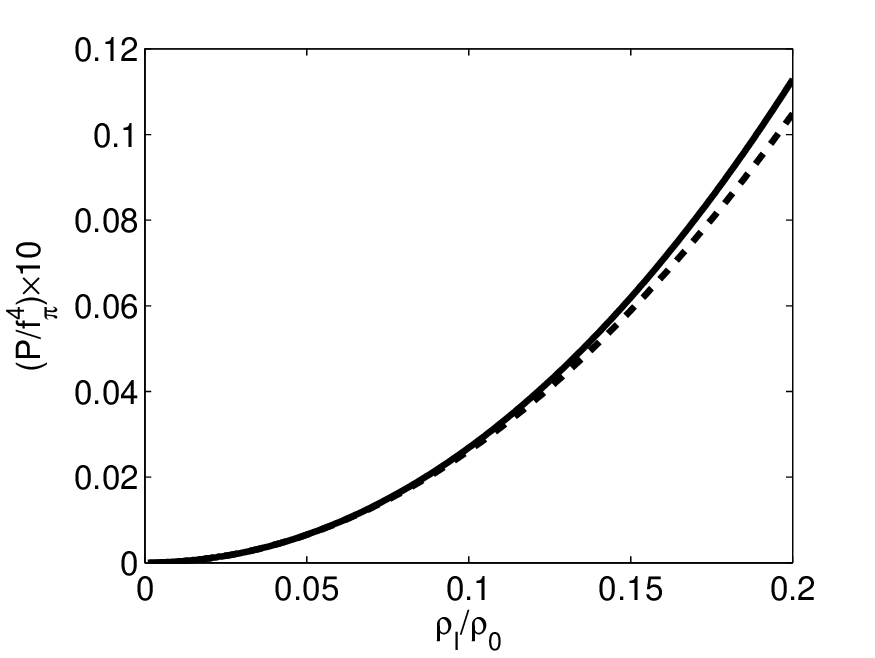}
\caption{The isospin chemical potential $\mu_{\text I}$ and
pressure ${\cal P}$ at low isospin density, calculated with the
NJL model and LHY theory.} \label{fig1}
\end{center}
\end{figure}

From the above gap and density equations, a nonzero isospin
density is associated with a nonzero pion condensate~\cite{njl4}.
In the limit of $\rho_{\text I}\rightarrow 0$, we have
analytically $\mu_{\text I}\rightarrow m_\pi$. In Fig.\ref{fig1},
we show the numerically calculated isospin chemical potential
$\mu_{\text I}$ and pressure ${\cal P}=-\Omega$ in the NJL model
and compare them to the LHY result (\ref{lhy2}) with the standard
Weinberg value $a=m_\pi/(16\pi f_\pi^2)$. Note that the higher
order correction $O(\sqrt{\rho_{\text I}a^3})$ in LHY is too small
to be observed when the isospin density is not high enough. We
find a very good agreement of the two calculations at sufficiently
low density. At $\rho_{\text I}\sim 0.1\rho_0$, the composite
nature of pions emerges and the NJL model starts to deviate from
the LHY result. Note that for baryon matter, the NJL model agrees
with the low-density result (\ref{baryon}) only for some special
model parameters.
\begin{figure}[!htb]
\begin{center}
\includegraphics[width=7cm]{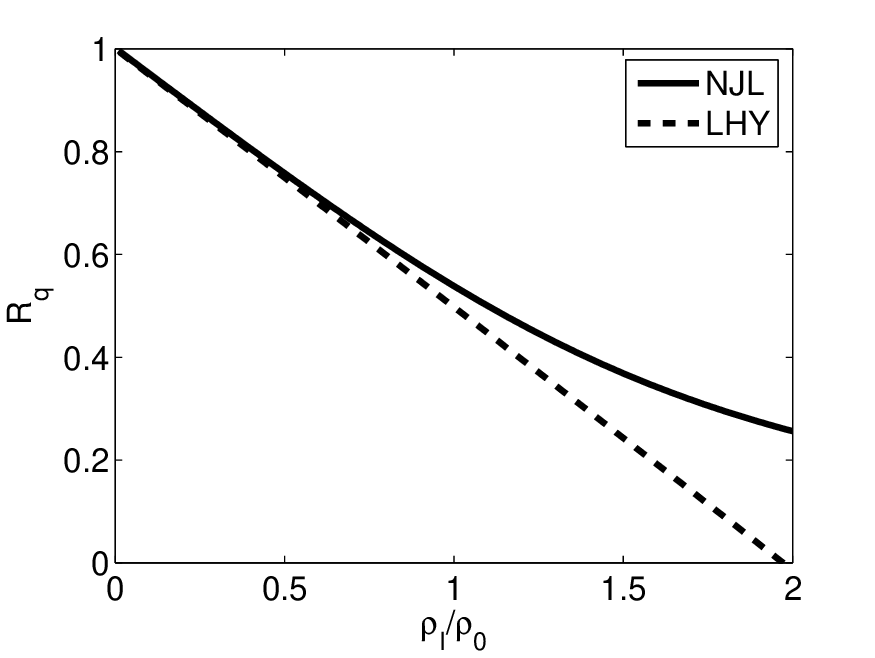}
\includegraphics[width=7cm]{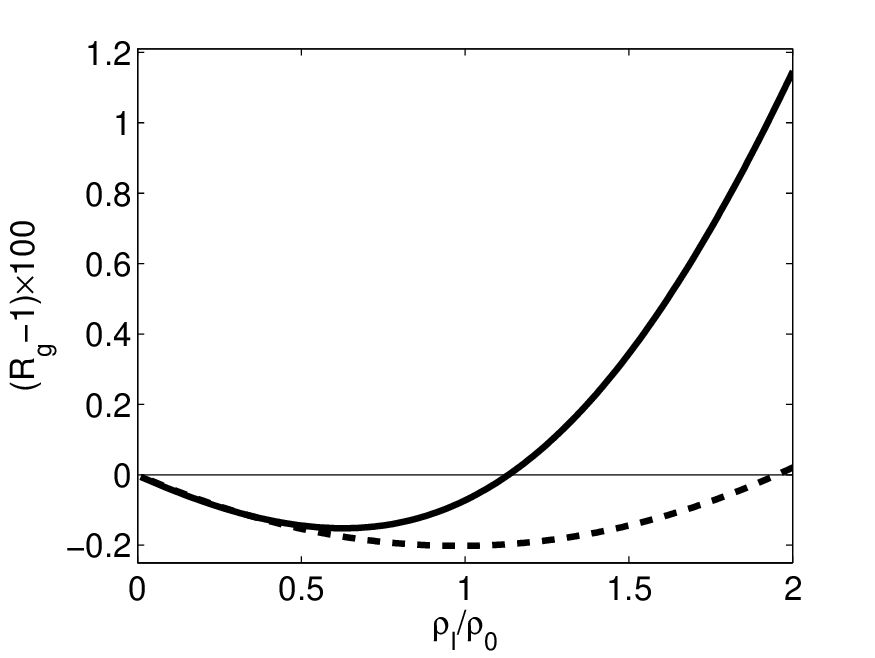}
\caption{The condensate ratios $R_q=\langle \bar
qq\rangle_{\rho_{\text I}}/\langle\bar qq\rangle_0$ and
$R_g=\langle G^2\rangle_{\rho_{\text I}}/\langle G^2\rangle_0$
from low to intermediate isospin density, calculated in the NJL
model and LHY theory with $\langle G^2\rangle_0=(360\ {\text
{MeV}})^4$.} \label{fig2}
\end{center}
\end{figure}

Having the correct low density limit, we then consider the quark
and gluon condensates at finite isospin density. The quark
condensate is obtained via either directly solving the NJL gap
equations (\ref{gap}) or the Hellmann-Feynman result (\ref{qc})
through the NJL energy density ${\cal E}=-{\cal P}+\mu_{\text
I}\rho_{\text I}$, and the gluon condensate can be calculated via
Eq. (\ref{gc}) together with the Hellmann-Feynman result
(\ref{qc}). The numerical results from the LHY theory and NJL
model are shown in Fig.\ref{fig2} for $\rho_{\text I}$ up to
$2\rho_0$. At very low density, the NJL model agrees well with the
model independent result, but at intermediate density, the LHY
theory for the dilute pion gas fails and the difference between
the two results becomes significant. The quark condensate
decreases much faster than the gluon condensate. At $\rho_{\text
I}=\rho_0$, the change in the gluon condensate is very small, but
the quark condensate has already been reduced by about $45\%$,
which is almost the same as the condensate suppression in baryon
matter at nuclear saturation density. The other significant
characteristic of the gluon condensate is its non-monotonous
density dependence. While this property is outside the validity
density region of the LHY theory, it is confirmed in the NJL
model. The gluon condensate firstly drops down, then turns to go
up at $\rho_{\text I}\simeq0.6\rho_0$ in the NJL model and
$\rho_{\text I} \simeq \rho_0$ in the LHY theory, and finally
becomes larger than its vacuum value at $\rho_{\text
I}>1.14\rho_0$ and $\rho_{\text I}>2\rho_0$ in the two
calculations.

The strong suppression of the quark condensate and the enhancement
of gluon condensate at intermediate density will induce the
evolution from a weakly interacting Bose condensate to a strongly
interacting Fermi superfluid, i.e., the so-called BEC-BCS
crossover~\cite{bcsbec}. From the fermion excitation spectrum, a
Fermi surface is opened with $\mu_{\text I}/2>M$~\cite{bcsbecq} at
$\rho_{\text I}>1.6\rho_0$. This type of strongly interacting
Fermi superfluid may be a realization of the quarkyonic matter
proposed recently~\cite{quarkyonic}. In such a matter, the chiral
symmetry is approximately restored, but the quarks are still
confined. Our prediction on the non-monotonous gluon condensate in
isospin matter is consistent with the expectation that the hadron
and quark phases are continued and there may exist no
deconfinement phase transition at zero temperature~\cite{lattice}.

In summary, we have investigated the quark and gluon condensates
in isospin matter. At low isospin density, we derived the
model-independent condensates by taking the Hellmann-Feynman
theorem for a dilute pion gas. Unlike the baryon matter, the
condensates in isospin matter are determined by precise pion
properties. Both the low density formula and the NJL model
calculation show that the repulsive interactions between the
charged pions may induce an unusual behavior of the gluon
condensate: it decreases slightly at low density and turns to
increase at intermediate density. While there are no lattice data
of the gluon condensate at finite isospin density, we expect our
conclusion can be examined in the future lattice calculations.

Our calculation for isospin matter can be easily extended to QCD
at finite isospin density $\rho_{\text I}$ and strangeness density
$\rho_{\text S}$ which is associated with kaon condensation
~\cite{cpt1}. For kaon matter with $\rho_{\text S}=2\rho_{\text
I}$, taking into account the relation
$m_K^2f_K^2=-(m_s+m_q)(\langle\bar s s\rangle_0+\langle\bar q
q\rangle_0)/2$ for the strangeness condensate $\langle\bar s
s\rangle_0$ in vacuum and the kaon mass $m_K$ and kaon decay
constant $f_K$, while the light and strange quark condensates
behave differently, the gluon condensate in (\ref{qgc2}) for pion
matter is valid for kaon matter, if we replace $m_\pi,\
\rho_{\text I}$ and $a$ by $m_K$, $\rho_{\text S}$ and kaon-kaon
scattering length $a_K$ in $I=1$ channel, $D_g(\rho_{\text
S})=-8/9\left(m_K\rho_{\text S}/2-4\pi a_K\rho_{\text S}^2/m_K
+\cdots\right)$. Taking the recent lattice QCD result
$m_Ka_K=0.352$~\cite{kaon}, the turning density where the gluon
condensate starts to increase is $\rho_{\text S}\simeq f_K^2m_K$
for kaon matter which is the same as $\rho_{\text I}\simeq
f_\pi^2m_\pi$ for pion matter.

{\bf Acknowledgement:} The work is supported by the NSFC Grant
10735040 and the National Research Program Grants 2006CB921404 and
2007CB815000.

\end{document}